\documentclass[11pt]{article}
\usepackage{amsmath,amsfonts,latexsym,graphicx,amssymb}
\usepackage[dvips]{epsfig}
\usepackage{amsfonts}
\usepackage{bm}
\date{}

\newcommand{\ot}{{\,\otimes\,}}
\newcommand{{\Cd}}{{\mathbb{C}^d}}

\def\<{\langle}
\def\>{\rangle}
\newtheorem{theorem}{Theorem}
\newtheorem{corollary}{Corollary}
\newtheorem{Proposition}{Proposition}
\newtheorem{DEF}{Definition}
\newtheorem{remark}{Remark}

\numberwithin{equation}{section}

\begin{document}
\title{\bf Spectral conditions for positive maps} \author{Dariusz
Chru\'sci\'nski and Andrzej Kossakowski \\
Institute of Physics, Nicolaus Copernicus University,\\
Grudzi\c{a}dzka 5/7, 87--100 Toru\'n, Poland}

\maketitle

\begin{abstract}

We provide a partial classification of positive linear maps in
matrix algebras which is based on a family of spectral conditions.
This construction generalizes celebrated Choi example of a map which
is positive but not completely positive. It is shown how the
spectral conditions enable one to construct  linear maps on tensor
products of matrix algebras which are positive but only on a convex
subset of separable elements. Such maps provide basic tools to study
quantum entanglement in multipartite systems.

\end{abstract}

\maketitle

\section{Introduction}

One of the most important problems of quantum information theory
\cite{QIT} is the characterization of mixed states of composed
quantum systems. In particular it is of primary importance to test
whether a given quantum state exhibits quantum correlation, i.e.
whether it is separable or entangled. For low dimensional systems
there exists simple necessary and sufficient condition for
separability. The celebrated Peres-Horodecki criterium
\cite{Peres,PPT} states that a state of a bipartite system living in
$\mathbb{C}^2 \ot \mathbb{C}^2$ or $\mathbb{C}^2 \ot \mathbb{C}^3$
is separable iff its partial transpose is positive. Unfortunately,
for higher-dimensional systems there is no single {\it universal}
separability condition.

It turns out that the above problem may be reformulated in terms of
positive linear maps in operator algebras: a state $\rho$ in
$\mathcal{H}_1 \ot \mathcal{H}_2$ is separable iff $({\rm id} \ot
\varphi)\rho$ is positive for any positive map $\varphi$  which
sends positive operators on $\mathcal{H}_2$ into positive operators
on $\mathcal{H}_1$. Therefore, a classification of positive linear
maps between operator algebras $\mathcal{B}(\mathcal{H}_1)$ and
$\mathcal{B}(\mathcal{H}_2)$ is of primary importance.
Unfortunately, in spite of the considerable effort, the structure of
positive maps is rather poorly understood
\cite{Stormer1}--\cite{OSID-W}. Positive maps play important role
both in physics and mathematics providing generalization of
$*$-homomorphism, Jordan homomorphism and conditional expectation.
Normalized positive maps define an affine mapping between sets of
states of $\mathbb{C}^*$-algebras.

In the present paper we perform partial classification of positive
linear maps which is based on spectral conditions. Actually,
presented method enables one to construct  maps with a desired
degree of positivity --- so called $k$-positive maps with
$k=1,2,\ldots,d=\min\{{\rm dim}\, {\cal H}_1,{\rm dim}\, {\cal
H}_2\}$. Completely positive (CP) maps correspond to $d$-positive
maps, i.e. maps with the highest degree of positivity. These maps
are fully classified due to Stinespring theorem
\cite{Stinespring,Paulsen}. Now, any positive map which is not CP
can be written as $\varphi = \varphi_+ - \varphi_-\,$, with
$\varphi_\pm$ being CP maps. However, there is no general method to
recognize the positivity of $\varphi$ from $\varphi_+ -
\varphi_-\,$. We show that suitable spectral conditions satisfied by
a pair $(\varphi_+,\varphi_-)$ guarantee $k$-positivity of
$\varphi_+ - \varphi_-\,$. This construction generalizes celebrated
Choi example of a map which is $(d-1)$-positive but not CP
\cite{Choi1}.

From the physical point of view our method leads to partial
classification of entanglement witnesses. Recall, that en
entanglement witness is a Hermitian operator $W \in
\mathcal{B}(\mathcal{H}_1 \ot \mathcal{H}_2)$ which is not positive
but satisfies $(h_1 \ot h_2,W\, h_1 \ot h_2) \geq 0$ for any $h_i
\in {\cal H}_i$.

Interestingly, our construction may be easily generalized for
multipartite case, i.e. for constructing entanglement witnesses in
$\mathcal{B}(\mathcal{H}_1 \ot \ldots \ot \mathcal{H}_n)$.
Translated into language of linear maps from
$\mathcal{B}(\mathcal{H}_2 \ot \ldots \ot \mathcal{H}_n)$ into
$\mathcal{B}(\mathcal{H}_1)$ presented method enables one to
construct maps which are not positive but which are positive when
restricted to separable elements in $\mathcal{B}(\mathcal{H}_2 \ot
\ldots \ot \mathcal{H}_n)$. To the best of our knowledge we provide
the first nontrivial example of such a map (nontrivial means that it
is not a tensor product of positive maps).

\section{Preliminaries}

Consider a space $\mathcal{L}(\mathcal{H}_1,\mathcal{H}_2)$ of
linear operators $a : \mathcal{H}_1 \longrightarrow \mathcal{H}_2$,
or equivalently a space of $d_1 \times d_2$ matrices, where $d_i =
{\rm dim}\, {\cal H}_i < \infty$. Let us recall that
$\mathcal{L}(\mathcal{H}_1,\mathcal{H}_2)$ is equipped with a family
of Ky Fan $k$-norms \cite{Horn}: for any $a \in
\mathcal{L}(\mathcal{H}_1,\mathcal{H}_2)$ one defines
\begin{equation}\label{}
    ||\, a\, ||_k := \sum_{i=1}^k s_i(a)\ ,
\end{equation}
where $s_1(a) \geq \ldots \geq s_d(a)$ ($d=\min\{d_1,d_2\}$) are
singular values of $a$. Clearly, for $k=1$ one recovers an operator
norm $||\, a \, ||_1 = ||\, a \, ||$ and if $d_1=d_2=d$, then for
$k=d$ one reproduces a trace norm $||\, a \, ||_d = ||\, a\, ||_{\rm
tr}$. The family of $k$-norms satisfies:

\begin{enumerate}

\item $ \ ||\, a\, ||_{k} \leq ||\, a\, ||_{k+1}\ ,$

\item $ \ ||\, a\, ||_{k} = ||\, a\, ||_{k+1}\ $ if and only if $\ {\rm rank}\,
a = k\ ,$

\item $ \, $ if $\,  {\rm rank}\, a \geq k+1\ ,$ then $\ ||\, a\, ||_{k} < ||\, a\, ||_{k+1}\ .$

\end{enumerate}
Note, that a family of Ky Fan norms may be equivalently introduced
as follows: let us define the following subset of ${\cal B}({\cal
H})$
\begin{equation}\label{}
    \mathcal{P}_k({\cal H}) = \{ \, p\in {\cal B}({\cal H}) : \ p=p^*=p^2\ , \ {\rm tr}\, p
    = k \, \} \ .
\end{equation}
Now, for any  $p \in \mathcal{P}_k({\cal H}_2)$  define the
following inner product in
$\mathcal{L}(\mathcal{H}_1,\mathcal{H}_2)$
\begin{equation}\label{}
    \< a,b\>_p := {\rm tr}\, [(pa)^*(pb)] =  {\rm tr}\, (a^*pb) = {\rm tr}\,
    (pba^*) \ .
\end{equation}
It is easy to show that
\begin{equation}\label{}
    ||\, a\, ||^2_k = \max_{p\in \mathcal{P}_k({\cal H}_2)}\, \< a,a\>_p  = \max_{p\in \mathcal{P}_k({\cal H}_2)}\, {\rm tr}\,
    (paa^*) \ .
\end{equation}

\noindent Thought out the paper we shall consider only finite
dimensional Hilbert spaces. We denote by $M_d$ a space of $d \times
d$ complex matrices and $\mathbb{I}_d$ is a identity matrix from
$M_d$.

\begin{Proposition} \label{Pr-I}
For arbitrary projectors $P$ and $Q$ in $\cal H$
\begin{equation}\label{}
    ||\, QPQ\, || = ||\, PQP\, ||\ .
\end{equation}
\end{Proposition}

\vspace{.5cm}

\noindent {\bf Proof.} One obviously has
\begin{equation}\label{}
 ||\, QPQ\, || = ||\, QP(QP)^*\, || = ||\, (QP)^2\, ||\ ,
\end{equation}
and
\begin{equation}\label{}
 ||\, PQP\, || = ||\, PQ(PQ)^*\, || = ||\, (PQ)^2\, ||\ .
\end{equation}
Now, due to $||\, A^2\, || = ||\, A^{*2}\, || = ||\, A\, ||^2$ one
obtains
\begin{equation}\label{}
 ||\, (QP)^2\, || =  ||\, (QP)^{*2}\, || =  ||\, (PQ)^2\, ||\ ,
\end{equation}
which ends the proof. \hfill $\Box$

\vspace{.5cm}

\noindent Consider now a Hilbert space being a tensor product ${\cal
H}_1 \ot {\cal H}_2$. Let us observe that any rank-1 projector $P$
in ${\cal H}_1 \ot {\cal H}_2$ may be represented in the following
way
\begin{equation}\label{P-F}
    P = \sum_{i,j=1}^{d_1} e_{ij} \ot F e_{ij} F^*\ ,
\end{equation}
where $F : \mathcal{H}_1 \longrightarrow \mathcal{H}_2$ and ${\rm
tr}\, FF^*=1$. Moreover, $\{e_1,\ldots,e_{d_1}\}$ denotes an
arbitrary orthonormal basis in ${\cal H}_1$, and $e_{ij} :=
|e_i\>\<e_j| \in \mathcal{B}(\mathcal{H}_1)$. Note, that $P =
|\psi\>\<\psi|$, where
\begin{equation}\label{}
    \psi = \sum_{i=1}^{d_1} e_i \ot Fe_i\ .
\end{equation}
It is easy to see that
\begin{equation}\label{}
    {\rm SR}(\psi) = {\rm rank}\, F\ ,
\end{equation}
where ${\rm SR}(\psi)$ denotes the Schmidt rank of $\psi$ ($1 \leq
{\rm SR}(\psi) \leq d)$, i.e. the number of non-vanishing Schmidt
coefficients in the Schmidt decomposition of $\psi$. It is clear
that $F$ does depend upon the chosen basis $\{
e_1,\ldots,e_{d_1}\}$. Note, however, that $FF^*$ is
basis-independent and, therefore, it has physical meaning being a
reduction of $P$ with respect to the first subsystem,
\begin{equation}\label{}
    FF^* = {\rm tr}_1 P\ .
\end{equation}

\begin{Proposition} \label{Pr-II}
Let $P$ be a projector in ${\cal H}_1 \ot {\cal H}_2$ represented as
in (\ref{P-F}) and $Q=\mathbb{I}_{d_1} \ot p$, where $p \in
\mathcal{P}_k({\cal H}_2)$. Then the following formula holds
\begin{equation}\label{}
    ||\,(\mathbb{I}_{d_1} \ot p) P (\mathbb{I}_{d_1} \ot p) \, ||  = {\rm tr} (pFF^*)\
    ,
\end{equation}
and hence
\begin{equation}\label{}
 ||\,(\mathbb{I}_{d_1} \ot p) P (\mathbb{I}_{d_1} \ot p) \, || \leq ||\, F\,
 ||^2_k \ .
\end{equation}

\end{Proposition}

\vspace{.5cm}

\noindent {\bf Proof.} Due to Proposition \ref{Pr-I} one has
\begin{equation}\label{}
||\, (\mathbb{I}_{d_1} \ot p) P (\mathbb{I}_{d_1} \ot p) \, || =
||\, P(\mathbb{I}_{d_1} \ot  p)P\, ||\ ,
\end{equation}
and hence
\begin{equation}\label{}
||\, (\mathbb{I}_{d_1} \ot p) P (\mathbb{I}_{d_1} \ot p) \, || =
{\rm tr}[P(\mathbb{I}_{d_1} \ot  p)] = \sum_{i=1}^{d_1} {\rm
tr}(Fe_{ii}F^*p) = {\rm tr}(FF^*p)\ ,
\end{equation}
where we have used $\sum_{i=1}^{d_1} e_{ii} = \mathbb{I}_{d_1}$.
\hfill $\Box$

\vspace{.5cm}

\noindent  Note, that if $F = V/\sqrt{d_1}$, where $V$ is an
isometry $VV^* = \mathbb{I}_{d_2}$, then $P$ is a maximally
entangled state
\begin{equation}\label{}
     P = \frac{1}{d_1} \sum_{i,j=1}^{d_1} e_{ij} \ot V e_{ij} V^*\ ,
\end{equation}
and one obtains in this case
\begin{equation}\label{}
||\,(\mathbb{I}_{d_1} \ot p) P (\mathbb{I}_{d_1} \ot p) \, || =
\frac{k}{d_1} = || \, F\, ||^2_k \ .
\end{equation}

\section{Entangled states vs. positive maps}

Let us recall that a state of a quantum system living in
$\mathcal{H}_1 \ot \mathcal{H}_2$ is separable iff the corresponding
density operator $\sigma$ is a convex combination of product states
$\sigma_1\ot \sigma_2$. For any normalized positive operator
$\sigma$ on $\mathcal{H}_1 \ot \mathcal{H}_2$ one may define its
Schmidt number
\begin{equation}\label{SN-rho}
    \mbox{SN}(\sigma) = \min_{\alpha_k,\psi_k}\, \left\{ \,
    \max_{k}\, \mbox{SR}(\psi_k)\, \right\}  \ ,
\end{equation}
where the minimum is taken over all possible pure states
decompositions
\begin{equation}\label{}
    \sigma = \sum_k \, \alpha_k\, |\psi_k\>\<\psi_k|\ ,
\end{equation}
with $\alpha_k\geq 0$, $\sum_k\, \alpha_k =1$ and $\psi_k$ are
normalized vectors in $\mathcal{H}_1 \ot \mathcal{H}_2$.  This
number characterizes the minimum Schmidt rank of the pure states
that are needed to construct such density matrix. It is evident that
$ 1 \leq \mbox{SN}(\sigma) \leq d=\min\{d_1,d_2\} $. Moreover,
$\sigma$ is separable iff $\mbox{SN}(\sigma) =1 $. It was proved
\cite{SN} that the Schmidt number is non-increasing under local
operations and classical communication. Now, the notion of the
Schmidt number enables one to introduce a natural family of convex
cones in $\mathcal{B}(\mathcal{H}_1 \ot \mathcal{H}_2)^+$ (a set of
semi-positive elements in $\mathcal{B}(\mathcal{H}_1 \ot
\mathcal{H}_2)$):
\begin{equation}\label{}
    \mathbf{V}_r = \{\, \sigma \in \mathcal{B}(\mathcal{H}_1 \ot \mathcal{H}_2)^+\ |\
    \mathrm{SN}(\sigma) \leq r\, \}  \ .
\end{equation}
One has the following chain of inclusions
\begin{equation}\label{V-k}
\mathbf{V}_1  \subset \ldots \subset  \mathbf{V}_d =
\mathcal{B}(\mathcal{H}_1 \ot \mathcal{H}_2)^+\ .
\end{equation}
 Clearly, $\mathbf{V}_1$ is a cone of
separable (unnormalized) states and $\mathbf{V}_d \smallsetminus
\mathbf{V}_1$ stands for a set of entangled states.

Let $\varphi : \mathcal{B}(\mathcal{H}_1) \longrightarrow
\mathcal{B}(\mathcal{H}_2)$ be a linear map such that $\varphi(a)^*
= \varphi(a^*)$. A map $\varphi$ is positive iff $\varphi(a) \geq 0$
for any $a \geq 0$.
\begin{DEF}
A linear map $\varphi$ is $k$-positive if
\[   {\rm id}_k\, \ot\, \varphi \ : \ M_k \ot \mathcal{B}(\mathcal{H}_1)\
\longrightarrow\ M_k \ot \mathcal{B}(\mathcal{H}_2)\ , \] is
positive. A map which is $k$-positive for $k=1,\ldots,d={\rm
min}\{d_1,d_2\}$ is called completely positive (CP map).
\end{DEF}
 Due to the Choi-Jamio{\l}kowski isomorphism \cite{Choi1,Jam} any
linear adjoint-preserving  map $\varphi : \mathcal{B}(\mathcal{H}_1)
\longrightarrow \mathcal{B}(\mathcal{H}_2)$ corresponds to a
Hermitian operator $\widehat{\varphi} \in \mathcal{B}(\mathcal{H}_1
\ot {\cal H}_2)$
\begin{equation}\label{}
\widehat{\varphi} :=  \sum_{i,j=1}^{d_1} e_{ij} \ot \varphi(e_{ij})\
.
\end{equation}

\begin{Proposition}
A linear map $\varphi$ is $k$-positive if and only if
\begin{equation}\label{}
    (\mathbb{I}_{d_1} \ot p)\widehat{\varphi}(\mathbb{I}_{d_1} \ot p) \geq 0
    \ ,
\end{equation}
for all $p \in \mathcal{P}_k({\cal H}_2)$. Equivalently, $\varphi$
is $k$-positive iff $\ {\rm tr}(\sigma \widehat{\varphi})\geq 0 $
for any $\sigma \in \mathbf{V}_k$.
\end{Proposition}

\begin{corollary}
A linear map $\varphi$ is positive iff $\ {\rm tr}(\sigma
\widehat{\varphi})\geq 0 $ for any $\sigma \in \mathbf{V}_1$, i.e.
or all separable states $\sigma$. Moreover, $\varphi$ is CP iff $\
{\rm tr}(\sigma \widehat{\varphi})\geq 0 $ for any $\sigma \in
\mathbf{V}_d$, i.e. $\widehat{\varphi} \geq 0$.
\end{corollary}

\section{Main result}

 It is well known that any CP map may be represented in the so
called Kraus form \cite{Kraus}
\begin{equation}\label{}
    \varphi_{\rm CP}(a) = \sum_\alpha K_\alpha a K^*_\alpha\ ,
\end{equation}
where (Kraus operators) $K_\alpha \in
\mathcal{L}(\mathcal{H}_1,\mathcal{H}_2)$. Any positive map is a
difference of two CP maps $\varphi = \varphi_+ - \varphi_-$.
However, there is no general method to recognize the positivity of
$\varphi$ from $\varphi_+ - \varphi_-$.  Consider now a special
class when $\widehat{\varphi}_+$ and $\widehat{\varphi}_-$ are
orthogonally supported and $\widehat{\varphi}_- = \lambda_1 P_1$,
with $P_1$ being a rank-1 projector. Let
\begin{equation}\label{map-1}
    \varphi(a) =  \sum_{\alpha=2}^{D} \lambda_\alpha F_\alpha a
    F_\alpha^* - \lambda_1 F_1 a F_1^*\ ,
\end{equation}
such that
\begin{enumerate}
\item all rank-1 projectors $P_\alpha = d_1^{-1} \sum_{i,j=1}^{d_1}
e_{ij} \ot F_\alpha e_{ij} F_\alpha^*$, are mutually orthogonal,

\item $\lambda_\alpha >0\ $, for $\alpha = 1,\ldots,D$, with $D:=d_1d_2$.
\end{enumerate}

\begin{theorem}   \label{Main-I}
Let $||\, F_1\, ||_k < 1$. If
\begin{equation}\label{M1}
\widehat{\varphi}_+\, \geq\, \frac{\lambda_1\, || F_1 ||^2_k}{1 - ||
F_1 ||^2_k}\, (\mathbb{I}_{d_1} \ot \mathbb{I}_{d_2} - P_1)\ ,
\end{equation}
then $\varphi$  is $k$-positive.
\end{theorem}

\noindent {\bf Proof.} Let $p \in \mathcal{P}_k(\mathcal{H}_2)$.
Take a unit vector $\xi \in (\mathbb{I}_{d_1} \ot p)\mathbb{C}^{d_1}
\ot \mathbb{C}^{d_2}$ and set
\begin{equation}\label{mu}
    \mu = \frac{\lambda_1\, || F_1 ||^2_k}{1 - || F_1
||^2_k}\ .
\end{equation}
One obtains
\begin{equation}\label{}
    (\xi,(\mathbb{I}_{d_1} \ot p)\widehat{\varphi}(\mathbb{I}_{d_1} \ot
    p)\xi)  \geq \mu - (\mu + \lambda_1)(\xi,(\mathbb{I}_{d_1} \ot p)P_1(\mathbb{I}_{d_1} \ot
    p)\xi) \ .
\end{equation}
Now, using Proposition \ref{Pr-II} one has
\begin{equation}\label{}
(\xi,(\mathbb{I}_{d_1} \ot p)P_1(\mathbb{I}_{d_1} \ot
    p)\xi) \leq ||\,(\mathbb{I}_{d_1} \ot p)P_1(\mathbb{I}_{d_1} \ot
    p)\, || \leq || F_1||^2_k\ ,
\end{equation}
and hence
\begin{equation}\label{}
(\xi,(\mathbb{I}_{d_1} \ot p)\widehat{\varphi}(\mathbb{I}_{d_1} \ot
    p)\xi) \geq 0 \ ,
\end{equation}
which proves $k$-positivity of $\varphi$. \hfill $\Box$

\begin{remark} {\em
Note, that condition (\ref{M1})  may be equivalently rewritten as
follows
\begin{equation}\label{M1-prim}
    \lambda_\alpha \geq \mu \ ; \ \ \ \alpha=2,\ldots,D\ ,
\end{equation}
with $\mu$ defined in (\ref{mu}).  }
\end{remark}

\begin{remark} {\em
If $d_1=d_2=d$ and $P_1$ is a maximally entangled state in
$\mathbb{C}^d \ot \mathbb{C}^d$, i.e. $F = U/\sqrt{d}$ with unitary
$U$, then the above theorem reproduces 25 years old result by
Takasaki and Tomiyama \cite{TT}. }
\end{remark}

\begin{remark} {\em
For $d_1=d_2=d\,$, $k=1$ and arbitrary $P_1$ the formula
(\ref{M1-prim}) was derived  by Benatti et. al. \cite{Benatti}. }
\end{remark}

The above theorem may be easily generalized for maps where ${\rm
rank}\, \widehat{\varphi}_- = m > 1$. Consider
\begin{equation}\label{general}
    \varphi(a) = \sum_{\alpha=m+1}^{D} \lambda_\alpha F_\alpha a
    F^*_\alpha - \sum_{\alpha=1}^m \lambda_\alpha F_\alpha a
    F^*_\alpha \ ,
\end{equation}
with $\lambda_\alpha >0$.
\begin{theorem}   \label{Main-II}
Let $\sum_{\alpha=1}^m||\, F_\alpha\, ||^2_k < 1$. If
\begin{equation}\label{}
\widehat{\varphi}_+ \ \geq \ \frac{\sum_{\alpha=1}^m \lambda_\alpha
|| F_\alpha ||^2_k}{1 - \sum_{\alpha=1}^m|| F_\alpha ||^2_k}\,
\left(\mathbb{I}_{d_1} \ot \mathbb{I}_{d_2} - \sum_{\alpha=1}^m
P_\alpha \right)\ ,
\end{equation}
then $\varphi$  is $k$-positive.
\end{theorem}
The proof is analogous.

\begin{remark} {\em
Note, that condition (\ref{M1})  may be equivalently rewritten as
follows
\begin{equation}\label{}
    \lambda_\alpha \geq \nu \ ; \ \ \ \alpha=m+1,\ldots,D\ ,
\end{equation}
with $\nu$ defined by
\begin{equation}\label{}
    \nu = \frac{\sum_{\alpha=1}^m \lambda_\alpha ||
F_\alpha ||^2_k}{1 - \sum_{\alpha=1}^m|| F_\alpha ||^2_k}\ .
\end{equation}
}
\end{remark}
Let us note that the condition $\lambda_\alpha > 0$ may be easily
relaxed. One has the following
\begin{corollary}
Consider a map (\ref{general}) such that $\lambda_1 = \ldots =
\lambda_\ell = 0$ ($\ell < m$) and $\lambda_{\ell +
1},\ldots,\lambda_{D}
> 0$. If
\begin{equation}\label{}
\widehat{\varphi}_+\ \geq\ \frac{\sum_{\alpha=\ell}^m \lambda_\alpha
|| F_\alpha ||^2_k}{1 - \sum_{\alpha=1}^m|| F_\alpha ||^2_k}\,
\left(\mathbb{I}_{d_1} \ot \mathbb{I}_{d_2} - \sum_{\alpha=1}^m
P_\alpha \right)\ ,
\end{equation}
then $\varphi$  is $k$-positive.
\end{corollary}
Consider again the map (\ref{map-1}).

\begin{theorem}   \label{Main-III}
Let $||\, F_1\, ||_k < 1$. If
\begin{equation}\label{M3}
\widehat{\varphi}_+\, < \, \frac{\lambda_1\, || F_1 ||^2_k}{1 - ||
F_1 ||^2_k}\, (\mathbb{I}_{d_1} \ot \mathbb{I}_{d_2} - P_1)\ ,
\end{equation}
then $\varphi$  is not $k$-positive.
\end{theorem}

\vspace{.5cm}

\noindent {\bf Proof.} To prove that $\varphi$ is not $k$ positive
we construct a vector $\xi_0 \in \mathbb{C}^{d_1} \ot
\mathbb{C}^{d_2}$ such that
\begin{equation}\label{}
    (\xi_0,(\mathbb{I}_{d_1} \ot p_0)\widehat{\varphi}(\mathbb{I}_{d_1} \ot
    p_0)\xi_0) < 0 \ ,
\end{equation}
for some $p_0 \in \mathcal{P}_k(\mathbb{C}^{d_2})$. Now, take any $p
\in \mathcal{P}_k(\mathbb{C}^{d_2})$ such that
\begin{equation}\label{}
    N^2 = {\rm tr}(pF_1F_1^*) \ ,
\end{equation}
is finite. Define
\begin{equation}\label{}
    \xi = N^{-1} \sum_{i=1}^{d_1} e_i \ot pF_1 e_i \ .
\end{equation}
Assuming (\ref{M3}) one finds
\begin{eqnarray}\label{}
(\xi,(\mathbb{I}_{d_1} \ot p)\widehat{\varphi}(\mathbb{I}_{d_1} \ot
    p)\xi) &<& \mu - (\mu + \lambda_1)(\xi,(\mathbb{I}_{d_1} \ot p)P_1(\mathbb{I}_{d_1} \ot
    p)\xi)\nonumber \\ &=& \frac{\mu}{|| F_1||^2_k} \, \Big[
    ||F_1||^2_k - (\xi,(\mathbb{I}_{d_1} \ot p)P_1(\mathbb{I}_{d_1} \ot
    p)\xi) \Big] \ ,
\end{eqnarray}
with $\mu$ defined by (\ref{mu}). Now, it is easy to show that
\begin{equation}\label{}
(\xi,(\mathbb{I}_{d_1} \ot p)P_1(\mathbb{I}_{d_1} \ot
    p)\xi) = {\rm tr}(pF_1F_1^*)\  ,
\end{equation}
and therefore
\begin{equation}\label{}
(\xi,(\mathbb{I}_{d_1} \ot p)\widehat{\varphi}(\mathbb{I}_{d_1} \ot
    p)\xi) < \frac{\mu}{|| F_1||^2_k} \, \Big[
    ||F_1||^2_k - {\rm tr}(pF_1F_1^*) \Big] \ .
\end{equation}
Finally, let us observe that since $\mathcal{P}_k(\mathbb{C}^{d_2})$
is compact there exists a point $p_0 \in
\mathcal{P}_k(\mathbb{C}^{d_2})$ such that
\begin{equation}\label{}
    {\rm Tr}(p_0F_1F_1^*) = || F_1||^2_k \ .
\end{equation}
Hence
\begin{equation}\label{}
    (\xi_0,(\mathbb{I}_{d_1} \ot p_0)\widehat{\varphi}(\mathbb{I}_{d_1} \ot
    p_0)\xi_0) < 0 \ ,
\end{equation}
with $\xi_0 = || F_1||^{-1}_k \sum_{i=1}^{d_1} e_i \ot p_0 F_1 e_i$.
\hfill $\Box$

\begin{corollary} Let $|| F_1 ||_{k+1} < 1$.
A map (\ref{map-1}) is $k$-positive but not $(k+1)$-positive if
\begin{equation}\label{}
\frac{\lambda_1\, || F_1 ||^2_{k+1}}{1 - || F_1 ||^2_{k+1}}\,
(\mathbb{I}_{d_1} \ot \mathbb{I}_{d_2} - P_1) \  > \
\widehat{\varphi}_+\ \geq \ \frac{\lambda_1\, || F_1 ||^2_k}{1 - ||
F_1 ||^2_k}\, (\mathbb{I}_{d_1} \ot \mathbb{I}_{d_2} - P_1)\ .
\end{equation}
\end{corollary}

\section{Example: generalized Choi maps}

Let us consider a family of maps
\[    \varphi_\lambda \ : \ M_d \ \longrightarrow\ M_d\ ,  \]
defined as follows
\begin{equation}\label{E}
    \varphi_\lambda(a) :=  \mathbb{I}_d \, {\rm tr} a  - \lambda F_1 a F_1^*\ .
\end{equation}
It generalizes celebrated  Choi  map which is $(d-1)$-positive but
not CP
\begin{equation}\label{}
    \varphi_{\rm Choi}(a) :=  \mathbb{I}_d \, {\rm tr} a  -  \frac{d}{d-1} a \
    ,
\end{equation}
which follows from (\ref{E}) with $F_1 = \mathbb{I}_d /\sqrt{d}$ and
$\lambda = d/(d-1)$. If $\lambda=d$, then (\ref{E}) reproduces the
so called reduction  map
\begin{equation}\label{}
 \varphi_{\rm red}(a) :=  \mathbb{I}_d \, {\rm tr} a  -   a\ ,
\end{equation}
which is known to be completely co-positive. One easily finds
\begin{equation}\label{}
    \widehat{\varphi}_\lambda =  \mathbb{I}_d \ot \mathbb{I}_d - \lambda P_1 \ ,
\end{equation}
 where
\begin{equation}\label{}
    P_1 = \sum_{i,j=1}^d e_{ij} \ot F_1 e_{ij} F^*_1\ .
\end{equation}
Let $f_k := ||\, F_1\, ||_k$ and assume that $f_{k+1} < 1\,$. A map
$\varphi_\lambda$ is $k$-positive but  not $(k+1)$-positive iff
\begin{equation}\label{}
    \frac{1}{d\, f_{k}}\ \geq\  \lambda\ >\ \frac{1}{d\, f_{k+1}} \ .
\end{equation}
Consider a family of states
\begin{equation}\label{}
    \rho_\mu = \frac{1-\mu}{d^2-1}\,  (\mathbb{I}_d \ot \mathbb{I}_d - P_1) + \mu P_1 \  .
\end{equation}
 Computing ${\rm
tr}(\widehat{\varphi}_\lambda \rho_\mu)$ one finds that ${\rm
SN}(\rho_\mu)=k$ iff
\begin{equation}\label{}
    f_{k}\ \geq\ \mu\  >\ f_{k-1}\ .
\end{equation}
In particular $\rho_\mu$ is separable iff $\mu \geq f_1=||\, F_1\,
||^2$. Note, that if $P_1$ is a maximally entangled state then
$\rho_\mu$ defines a family of isotropic state. In this case $f_k =
k/d$ and one recovers well know result \cite{SN}: ${\rm
SN}(\rho_\mu)=k$ iff $ k/d \geq \mu > (k-1)/d$.

Consider now the following generalization of (\ref{E}):
\begin{equation}\label{E'}
    \varphi_\lambda(a) := \mathbb{I}_d \, {\rm tr} a  - \lambda \sum_{\alpha=1}^{m}F_\alpha a F_\alpha^*\
    ,
\end{equation}
and the corresponding operator
\begin{equation}\label{}
    \widehat{\varphi}_\lambda =  \mathbb{I}_d \ot \mathbb{I}_d - \lambda P \ ,
\end{equation}
 where $P$ is a rank-$m$ projector given by
\begin{equation}\label{}
    P = \sum_{i,j=1}^d \sum_{\alpha=1}^{m}\, e_{ij} \ot F_\alpha e_{ij} F^*_\alpha\ .
\end{equation}
 A map $\varphi_\lambda$ is
$k$-positive  if
\begin{equation}\label{}
  \lambda\ \leq\ \frac{1}{d\, \widetilde{f}_{k}} \ ,
\end{equation}
where now $\widetilde{f}_k = \sum_{\alpha=1}^{m-1}||\, F_\alpha\,
||^2_k\,$ and we assume that $\widetilde{f}_k < 1$. Consider a
family of states
\begin{equation}\label{}
    \rho_\mu = \frac{1-m\mu}{d^2-m}\ (\mathbb{I}_d \ot \mathbb{I}_d -P) + \frac{\mu}{m}\, P \  .
\end{equation}
 Computing ${\rm
tr}(\widehat{\varphi}_\lambda \rho_\mu)$ one finds that ${\rm
SN}(\rho_\mu)=k$ iff
\begin{equation}\label{}
    \widetilde{f}_{k}\ \geq\ \mu\  >\ \widetilde{f}_{k-1}\ .
\end{equation}
In particular $\rho_\mu$ is separable iff $\mu \geq
\widetilde{f}_1=\sum_{\alpha=1}^{m-1}||\, F_\alpha\, ||^2$. Note,
that if $P$ is a sum of $m$ maximally entangled state then
$\rho_\mu$ defines a generalization of a family of isotropic state.
In this case $\widetilde{f}_k = mk/d$ and one obtains: ${\rm
SN}(\rho_\mu)=k$ iff $ mk/d \geq \mu
> m(k-1)/d$.

\section{Multipartite setting}

Consider now an $n$-partite state $\rho$ living in $\mathcal{H}_1
\ot \ldots \ot \mathcal{H}_n$. Recall
\begin{DEF}
A state $\rho$ is separable iff it can be represented as the convex
combination of product states $\rho_1 \ot \ldots \ot \rho_n$.
\end{DEF}

\begin{theorem}
An $n$-partite  state $\rho$ in $\mathcal{H}_1 \ot \ldots \ot
\mathcal{H}_n$ is separable iff
\begin{equation}\label{}
    ({\rm id} \ot \varphi)\, \rho \geq 0 \ ,
\end{equation}
for all linear maps $\varphi : {\cal B}(\mathcal{H}_2 \ot \ldots \ot
\mathcal{H}_n) \longrightarrow {\cal B}({\cal H}_1)$ satisfying
\begin{equation}\label{PS}
    \varphi(p_2 \ot \ldots \ot p_n) \geq 0 \ ,
\end{equation}
where $p_k$ is a rank-1 projector in ${\cal H}_k$.
\end{theorem}

\begin{DEF}[Generalized Choi-Jamio{\l}kowski isomorphism]
For any linear map $$\varphi : {\cal B}(\mathcal{H}_2 \ot \ldots \ot
\mathcal{H}_n) \longrightarrow {\cal B}({\cal H}_1)\ ,$$ define an
operator $\widehat{\varphi}$ in ${\cal B}(\mathcal{H}_1 \ot \ldots
\ot \mathcal{H}_n)$
\begin{equation}\label{}
    \widehat{\varphi} := d_1({\rm id} \ot \varphi^\sharp)\, P^+ \ ,
\end{equation}
where $P^+$ is the canonical maximally entangled state in
$\mathcal{H}_1 \ot \mathcal{H}_1$, and $\varphi^\sharp$ denotes a
dual map.
\end{DEF}

\begin{Proposition}
A linear map $$\varphi : {\cal B}(\mathcal{H}_2 \ot \ldots \ot
\mathcal{H}_n) \longrightarrow {\cal B}({\cal H}_1)\ ,$$ satisfies
(\ref{PS}) iff
\begin{equation}\label{PS-prim}
    {\rm tr} [(p_1 \ot \ldots \ot p_n) \, \widehat{\varphi}] \geq 0 \ ,
\end{equation}
for any rank-1 projectors $p_k$.
\end{Proposition}

\vspace{.5cm}

\noindent {\bf Proof.} One has
\begin{equation}\label{}
 {\rm tr} [(p_1 \ot \ldots \ot p_n) \, \widehat{\varphi}] = d_1 {\rm tr} [(p_1 \ot \ldots \ot p_n) \, ({\rm id} \ot
 \varphi^\sharp) P^+] = d_1 {\rm tr} [P^+\cdot p_1 \ot \varphi(p_2 \ot \ldots
 \ot p_n)] \ .
\end{equation}
Now, using $P^+ = d_1^{-1} \sum_{i,j=1}^{d_1} e_{ij} \ot e_{ij}$ and
obtains
\begin{equation}\label{}
{\rm tr} [P^+ \cdot p_1 \ot \varphi(p_2 \ot \ldots
 \ot p_n)]  = d_1^{-1} \sum_{i,j=1}^{d_1} {\rm tr}(e_{ij}p_1) \, {\rm tr}[e_{ij} \varphi(p_2 \ot
 \ldots \ot  p_n)]\ .
\end{equation}
Finally, due to $\sum_{i,j}{\rm tr}(e_{ij}a) e_{ij} = a^T$, one
finds
\begin{equation}\label{}
 {\rm tr} [(p_1 \ot \ldots \ot p_n) \, \widehat{\varphi}] =
 {\rm tr}[p_1^T \varphi(p_2 \ot \ldots \ot
 p_n)]\ ,
\end{equation}
from which the Proposition immediately follows. \hfill $\Box$

\begin{corollary}
A linear map $$\varphi : {\cal B}(\mathcal{H}_2 \ot \ldots \ot
\mathcal{H}_n) \longrightarrow {\cal B}({\cal H}_1)\ ,$$ satisfies
(\ref{PS}) iff
\begin{equation}\label{PS-prim}
    (\mathbb{I} \ot p_2 \ot \ldots \ot p_n) \, \widehat{\varphi} \, (\mathbb{I} \ot p_2 \ot \ldots
    \ot p_n) \geq 0 \ ,
\end{equation}
for any rank-1 projectors $p_k$.
\end{corollary}

To construct linear maps which are positive on separable states let
us define the following norm: let
\begin{equation}\label{}
    \mathcal{P}_{\rm sep} = \{ p_2 \ot \ldots \ot p_n\ : \
    p_k=p^*_k=p^2_k\ , \ {\rm tr}\,p_k=1 \} \ ,
\end{equation}
and define an inner product in the space of linear operators
$\mathcal{L}(\mathcal{H}_1,\mathcal{H}_2 \ot \ldots \ot
\mathcal{H}_n)$
\begin{equation}\label{}
    \< A,B\>_P := {\rm tr}[(PA)^*(PB)]\ ,
\end{equation}
with $P \in  \mathcal{P}_{\rm sep}$. Finally, let
\begin{equation}\label{}
    ||\, A\, ||^2_{\rm sep} := \max_{P\in  \mathcal{P}_{\rm sep}}
    \<A,A\>_P\ .
\end{equation}
It is clear that
\begin{equation}\label{}
||\, A\, ||_{\rm sep} \leq ||\, A\, ||\ .
\end{equation}
Consider now a linear map defined by
\begin{equation}\label{}
    \varphi(a) = \sum_{\alpha=2}^{D} \lambda_\alpha F_\alpha a
    F^*_\alpha - \lambda_1 F_1 a F^*_1\ ,
\end{equation}
where $D=d_1\ldots d_n\,$, ${\rm tr}(F_\alpha^* F_\beta) =
\delta_{\alpha\beta}$ and $\lambda_\alpha > 0$. One finds for the
corresponding $\widehat{\varphi}$
\begin{equation}\label{}
\widehat{\varphi} = \sum_{\alpha=2}^{D} \lambda_\alpha P_\alpha -
\lambda_1 P_1\ ,
\end{equation}
where the rank-1 projectors read as follows
\begin{equation}\label{}
    P_\alpha = \sum_{i,j=1}^{d_1} e_{ij} \ot F_\alpha e_{ij}
    F^*_{\alpha}\ .
\end{equation}
In analogy to Theorems \ref{Main-II} and \ref{Main-III}  one easily
proves

\begin{theorem}   \label{Main-IV}
Let $||\, F_1\, ||_{\rm sep} < 1$. Then $\varphi$  is positive on
separable states if and only if
\begin{equation}\label{}
\lambda_\alpha\  \geq\  \frac{ \lambda_1 || F_1 ||_{\rm sep}^2}{1 -
|| F_1 ||^2_{\rm sep}}\ ,
\end{equation}
for $\alpha=2,\ldots,D$.
\end{theorem}

\begin{corollary}   \label{Main-V}
Let $||\, F_1\, ||_{\rm sep} < ||\, F_1\, || < 1$. Then $\varphi$ is
positive on separable states but not positive if and only if
\begin{equation}\label{}
\frac{ \lambda_1 || F_1 ||^2}{1 - || F_1 ||^2} \ > \ \lambda_\alpha\
\geq\ \frac{ \lambda_1 || F_1 ||_{\rm sep}^2}{1 - || F_1 ||^2_{\rm
sep}}\ ,
\end{equation}
for $\alpha=2,\ldots,D$.
\end{corollary}

\noindent {\bf Example.} Consider a map
\begin{equation}\label{}
    \varphi_\lambda \ : \ M_d \ot M_d \ \longrightarrow \ M_{d^2} \equiv M_d \ot M_d \ ,
\end{equation}
defined by
\begin{equation}\label{}
    \varphi_\lambda(a) = \lambda (\mathbb{I}_d \ot \mathbb{I}_d \, {\rm tr} a - F_0
    a F_0 ) -  F_0 a F_0\ ,
\end{equation}
with
\begin{equation}\label{}
    F_0 = F_0^* = \frac{1}{\sqrt{2d(d-1)}}\, \left[ \mathbb{I}_d \ot \mathbb{I}_d -
    \sum_{i,j=1}^d e_{ij} \ot e_{ij}^* \right]\ .
\end{equation}
Note that $ {\rm tr} F_0^2 = 1$ and $\sqrt{d(d-1)/2}\cdot F_0$ is a
projector (see \cite{channels,nasza} for more details). Hence
\begin{equation}\label{}
     ||\, F_0\, ||^2 = \frac{2}{d(d-1)}\ .
\end{equation}
Now, for any rank-1 projectors $p,q \in M_d$ one has
\begin{equation}\label{}
    {\rm tr}\Big[(p \ot q)F_0^2\Big] = \frac{1}{d(d-1)}\, (1 - {\rm tr}pq ) \
    ,
\end{equation}
and therefore
\begin{equation}\label{pqF0}
 ||\, F_0 ||^2_{\rm sep} :=   \max_{p,q\in \mathcal{P}_{\rm sep}} {\rm tr}\Big[(p \ot q)F_0^2\Big] =
    \frac{1}{d(d-1)} < ||\, F_0\, ||^2\ .
\end{equation}

\begin{corollary}
Let $d>2$, i.e. $||\, F_0\, ||_{\rm sep} < ||\, F_0\, || < 1$. For
\begin{equation}\label{}
\frac{2}{d(d-1)-2}\ > \     \lambda\ \geq \ \frac{1}{d(d-1)-1}
\end{equation}
$\varphi_\lambda$ is positive on separable elements in $M_d \ot M_d$
but it is not a positive map.
\end{corollary}

%\vspace{.2cm}

\begin{remark}
{\em To the best of our knowledge $\varphi_\lambda$ provides the
first nontrivial example of a map which is not positive but it is
positive on separable states. Nontrivial means that it is not a
tensor product of two positive maps. }
\end{remark}

\section{Conclusions}

We provide partial classification of positive linear maps based on
spectral conditions. Presented method generalizes celebrated Choi
example of a map which is positive but not CP. From the physical
point of view our scheme  provides simple method for constructing
entanglement witnesses. Moreover, this scheme may be easily
generalized for multipartite setting.

Presented method guarantees $k$-positivity but says nothing about
indecomposability and/or optimality. We stress that both
indecomposable and optimal positive maps are crucial in detecting
and classifying quantum entanglement. Therefore, the analysis of
positive maps based on spectral properties deserves further study.

\section*{Acknowledgement} This work was partially supported by the
Polish Ministry of Science and Higher Education Grant No
3004/B/H03/2007/33 and by the Polish Research Network  {\it
Laboratory of Physical Foundations of Information Processing}.


\begin{thebibliography}{1} \bibliographystyle{plain}

\bibitem{QIT} M. A. Nielsen and I. L. Chuang, {\it Quantum computation
and quantum information}, Cambridge University Press, Cambridge,
2000.


\bibitem{Peres} A. Peres, Phys. Rev. Lett. {\bf 77}, 1413 (1996).

\bibitem{PPT}  P. Horodecki, Phys. Lett. A {\bf 232}, 333 (1997).

%==============MAPS===============================

\bibitem{Stormer1}  E. St{\o}rmer, Acta Math. {\bf 110}, 233
(1963).

\bibitem{Arverson} W. Arverson, Acta Math. {\bf 123}, 141 (1969).

\bibitem{Choi1} M.-D. Choi,  Lin. Alg. Appl. {\bf 10}, 285 (1975);
{\em ibid} {\bf 12}, 95 (1975).

\bibitem{Choi2} M.-D. Choi, J. Operator Theory, {\bf 4}, 271
(1980).

\bibitem{Jam} A. Jamio{\l}kowski, Rep. Math. Phys. {\bf 3}, 275 (1972).

\bibitem{Woronowicz1} S.L. Woronowicz, Rep. Math. Phys. {\bf 10}, 165
(1976).

\bibitem{Woronowicz2} S.L. Woronowicz, Comm. Math. Phys. {\bf 51}, 243 (1976).

\bibitem{TT}  K. Takasaki and J. Tomiyama, Mathematische  Zeitschrift {\bf 184},
101-108 (1983).


\bibitem{Robertson1} A.G. Robertson, Quart. J. Math. Oxford (2),
{\bf 34}, 87 (1983)

\bibitem{Tang} W.-S. Tang, Lin. Alg. Appl. {\bf 79}, 33 (1986)


\bibitem{Osaka1} H. Osaka,  Lin. Alg. Appl. {\bf 153}, 73 (1991); {\em ibid} {\bf 186}, 45
(1993).

\bibitem{Osaka2} H. Osaka, Publ. RIMS Kyoto Univ. {\bf 28}, 747
(1992).

\bibitem{Cho-Kye} S. J. Cho, S.-H. Kye, and S.G. Lee, Lin. Alg.
Appl. {\bf 171}, 213 (1992).

\bibitem{Kye3} S.-H. Kye,  Lin. Alg. Appl. {\bf 362}, 57 (2003).


\bibitem{Ha1} K.-C. Ha,
Publ. RIMS, Kyoto Univ., {\bf 34}, 591 (1998).

\bibitem{Ha2} K.-C. Ha,  Lin. Alg.
Appl. {\bf 348}, 105 (2002); {\it ibid} {\bf 359}, 277 (2003).




\bibitem{Kossak1} A. Kossakowski, Open Sys. Information Dyn. {\bf 10},
213 (2003).

\bibitem{Benatti} F. Benatti, R. Floreanini and M. Piani,
Open Systems and Inf. Dynamics, {\bf 11}, 325-338 (2004).


\bibitem{Hall} W. Hall, J. Phys. A: Math. Gen. {\bf 39}, (2006)
14119.

\bibitem{Breuer} H.-P. Breuer, Phys. Rev. Lett. {\bf 97}, 0805001
(2006).

\bibitem{Ruskai} D. Perez-Garcia, M. M. Wolf, D. Petz and M. B.
Ruskai, J. Math. Phys. {\bf 47}, 083506 (2006).

\bibitem{atomic} D. Chru\'sci\'nski and A. Kossakowski, J. Phys. A:
Math. Theor.  {\bf 41}, 215201  (2008).

\bibitem{OSID-W} D. Chru\'sci\'nski and A. Kossakowski,  Open Systems and Inf.
Dynamics, {\bf 14}, 275 (2007).

%=================================================


\bibitem{Stinespring} W.F. Stinespring, Proc. Amer. Math. Soc.
{\bf 6}, 211 (1955).


\bibitem{Paulsen} V. Paulsen, {\it Completely Bounded Maps and Operator
Algebras}, Cambridge University Press, 2003.


\bibitem{Horn} R.A. Horn and C.R. Johnson, {\it Topics in Matrix Analysis},
 (Cambridge University Press, New York, 1991).

\bibitem{SN} B. Terhal and P. Horodecki,
Phys. Rev. A {\bf 61}, 040301 (2000)

\bibitem{Kraus} K. Kraus, {\it States, Effects and Operations: Fundamental Notions of Quantum Theory},
 Springer Verlag, 1983.

\bibitem{channels} D. Chru\'sci\'nski and A. Kossakowski, Open Systems and Inf.
Dynamics, {\bf 13}, 17-26 (2006).

\bibitem{nasza} D. Chru\'sci\'nski and A. Kossakowski, Phys.
Rev. A {\bf 73}, 062313 (2006).



\end{thebibliography}
\end{document}